\documentclass[useAMS,usenatbib]{mn2e}
\usepackage{amssymb}
\usepackage{amsmath}
\usepackage{graphicx}
\usepackage{BibDef}
\usepackage{hyperref}
\usepackage{subfigure}
\def\lsim{\mathrel{\rlap{\lower 3pt \hbox{$\sim$}} \raise 2.0pt \hbox{$<$}}}
\def\gsim{\mathrel{\rlap{\lower 3pt \hbox{$\sim$}} \raise 2.0pt \hbox{$>$}}}
\def\msun{\rm {M_{\large \odot}}}

\title[MBH dynamics: binary formation] {Massive black hole and gas dynamics in mergers of galaxy nuclei. II. Black hole sinking in star--forming 
nuclear discs.}
\author[Lupi et al.]{Alessandro Lupi$^{1}$, Francesco Haardt$^{1,2}$, Massimo Dotti$^{2,3}$ \& Monica Colpi$^{2,3}$.\\
$^1$DiSAT, Universit\`a degli Studi dell'Insubria, Via Valleggio 11, I-22100 Como, Italy\\
$^2$INFN, Sezione di Milano-Bicocca, Piazza della Scienza 3, I-20126 Milano, Italy\\
$^3$Dipartimento di Fisica, Universit\`a degli Studi di Milano-Bicocca, Piazza della Scienza 3, Milano I-20126, Italy}
\begin{document}

\date{Draft April 2015}

\pagerange{\pageref{firstpage}--\pageref{lastpage}} \pubyear{2015}

\maketitle

\label{firstpage}

\begin{abstract} 
Mergers of gas--rich galaxies are key events in the hierarchical
built--up of cosmic structures, and can lead to the formation of 
massive black hole binaries. By means of high--resolution hydrodynamical simulations we consider the
late stages of a gas--rich major merger, detailing the dynamics of two
circumnuclear discs, and of the hosted massive black holes during
their pairing phase.  During the merger gas clumps with masses of a
fraction of the black hole mass form because of fragmentation. Such
high--density gas is very effective in forming stars, and the most
massive clumps can substantially perturb the black hole orbits. After
$\sim 10$ Myr from the start of the merger a gravitationally bound
black hole binary forms at a separation of a few parsecs, and soon
after, the separation falls below our resolution limit of 0.39 pc. At
the time of binary formation the original discs are almost completely
disrupted because of SNa feedback, while on pc scales the residual gas
settles in a circumbinary disc with mass $\sim 10^5\,\msun$.  
We also test that binary dynamics is robust against the details of the SNa feedback employed in the simulations, while gas dynamics is not. We finally
highlight the importance of the SNa time--scale on our results.
\end{abstract}

\section{Introduction}
Understanding the black hole binary formation
path in galaxy's mergers is a key step towards understanding the mode of assembly of massive black holes (MBHs) across cosmic history \citep{dimatteo05, sesana14,dubois14b}.
The masses and spins of MBHs are sculpted by complex processes that involve episodes of accretion and mergers, during the hierarchical growth
of structures. The coalescence in a single, more massive BH (induced by gravitational wave emission) is expected to occur whenever the two MBHs
present in each of the interacting galaxies reach sub--parsec separation under the action of stellar and gas torques \citep{amaro-seoane13}. 

The dynamics of MBHs is complex as it occurs in the time--varying environment of a galaxy merger where stars, and cool gas, turning into new stars, evolve on comparable time--scales \citep[see][for a review]{colpi14}, and 
is customarily described as a three--step process: (I) {\it pairing} 
during the merger of the galaxies embedded in their dark matter haloes, resulting in the formation of a Keplerian binary, (II) {\it migration} by gas or/and {\it hardening} by stars ,
and lastly  (III) {\it gravitational wave driven inspiral}. 

The seminal paper by \citet{begelman80} showed that in an already established spherical galaxy remnant devoid of gas, the fastest of these three phases is that of pairing under the action of dynamical friction against stars. The MBHs are driven down to parsec separations and form a Keplerian binary, when 
the stellar mass enclosed in their orbit drops below their mass.
\citet{begelman80} further noticed that hardening of the binary by encounters with individual stars in phase (II) is the most critical and long--lived phase, representing a bottleneck 
to the path to coalescence, at least for MBHs with masses larger than
$\sim 2 \times 10^6 \msun$ \citep[see e.g.][]{merritt05,merritt07}. The so--called last parsec problem for the most massive MBHs
has been recently alleviated after considering more
realistic galaxy's backgrounds\footnote{We will not consider
here MBH pairing in pure stellar environments and refer to \citet{preto11,khan13,vasiliev14,vasiliev15} and \citet{sesana15} for recent findings.}. \citet{begelman80}  mention on the possibility
that gas can accelerate the orbital decay.  In this case, torques from
gas distributed around the binary in the form of a circumbinary disc are central for
driving the binary down to coalescence but they depend on how massive and long--lived are these discs and little is known about their formation, 
structure and lifetime in the relic galaxy \citep{cuadra09,roedig11,roedig12,delvalle12}.

Thanks to major advances in numerical computing, a number of studies have begun tracing  phase (I) of MBH pairing 
in the very early stages of a merger when the two dark matter haloes  first touch, and later when the galactic discs (stellar and gaseous) and bulges 
interact to drive and terminate the morphological transformation of the galaxies \citep{mayer07,roskar15}. A  leap in understanding the role of gas
during the pairing phase was taken when studying minor mergers, i.e. mergers with nominal 1:4  mass ratios and less.
It was demonstrated that thanks to its dissipative action, gas deepens the gravitational potential of 
the less massive galaxy reducing the action of tides by the primary \citep{kaza05}. This prevents the wandering to the lighter MBH in the periphery of the primary when sufficient gas is present in the host galaxy, raising the question as to whether minor mergers  lead in general to 
binary formation \citep{callegari09,callegari11}.  Follow--up studies have indicated that the dividing line from success and failure in forming an MBH binary (MBHB hereon) 
is around 1:10 mass ratios  \citep{bellovary10,vanwasse12}, but still depends on details such as the encounter geometry and gas content.

Major mergers among gas--rich galaxies represent a natural path for MBH pairing and binary formation. The orbital braking
is in these cases driven by gas--dynamical friction which is faster than dynamical friction from stars \citep{escala05,dotti06,mayer07,chapon13}.
Massive inflows of gas during
the merger lead to the formation of a circumnuclear disc comprising most of the gas initially hosted in the individual discs.  It is in this 
new disc that the two MBH
sink  forming a Keplerian system down to pc scale on a time typically of $\sim 5$ Myr after completion of the merger.
A critical moment was proven to be the time when the two discs start colliding and developing shocks due to orbit crossings 
 which cancel/redistribute
the gas angular momentum.  This process triggers inflows that lead to the formation of a turbulent Toomre stable, massive nuclear disc \citep{chapon13}.
The limit and drawback of these high--resolution simulations is that gas is treated as a single phase 
medium described by a polytropic equation of state (with index 7/5) which mimics the thermodynamics of a generic star--forming region. 

\citet{capelo15} studied the large--scale dynamics of MBH in a variety of mergers  with mass ratio 1:1 down to 1:10 
to explore the black hole accretion history and their dynamics during the pairing phase, down to separations of several parsecs.
Their focus was mainly in exploring the possibility of triggering `dual' AGN activity along the course of the merger.
Only recently state--of--the--art simulations of major mergers have achieved enough resolution to detail
the MBH dynamics down parsec scales, in presence of a multiphase gas \citep{roskar15}.

Three lines of studies have been developed in the past, showing that the MBH dynamics in star--forming regions is more complex than predicted by single--phase models and that the black hole environment is far richer than expected.  
The three approaches consist of: (a) simulations of major mergers of gas--rich disc galaxies starting from large scales \citep[$\gsim 100$ kpc;][]{roskar15}; (b)
simulations of two gaseous discs of mass $\sim 10^8\,\msun$ colliding on scales of $\sim 500$ pc, representing a zoom--in version of very gas--rich galaxy's discs in the verge of  merging \citep{lupi15}; (c) simulations of a single gaseous disc unstable to fragmentation
as it forms as a consequence of major gas-rich mergers. This last kind of
numerical experiment  focused on massive discs ($\lesssim 10^{8-9}\,\msun$) on kpc scale 
\citep{fiacconi13,delvalle15} as well as on lighter ($\lsim
10^7 \msun$) circumbinary discs \citep{amaroseoane13}\footnote{Noticeably,
the pairing of the MBHB in this last study has been followed using a high
accuracy direct $N$-body code to evolve the distribution of stars previously
obtained from an SPH run.}.

With the smoothed particle hydrodynamics (SPH) code \textsc{gasoline}, case (a) was explored by \citet{roskar15}  who simulated a prograde in--plane 1:1 merger of two late--type galaxies (Milky Way like)
each having a 10\% fraction of gas in their discs. Star formation and feedback were turned on at the onset
of the simulation to generate a multiphase medium. 5 Gyr after the start of the simulation, particle--splitting of the baryonic particles  was performed  in an excised zoom--in region to follow  the last 100 Myr of evolution when the galaxy's cores touch on scales of $\sim 5$ kpc.
The starburst triggered by the merger and stellar feedback create a nuclear region temporarily devoid of gas and only after 
$\sim 10$ Myr a  nuclear disc is rebuilt. The nature of the multiphase gas which develops clumps affects the MBH dynamics.
The MBHs undergo gravitational encounters with massive gas clouds and stochastic torquing
by both clouds and spiral modes in the disc relents the pairing process. The MBHs are kicked out of the plane due 
to their interaction with clumps and this delays the time of binary formation which now is $\sim 100$ Myr, 
about two orders of magnitude longer than in the idealised mergers with one--component gas.
This finding is in line with that found by \citet{fiacconi13} who simulated the dynamics of MBHs in a massive gaseous disc of $\lesssim 10^{8-9}\,\msun$ subject to rapid
cooling and fragmentation, in a suite of runs of type (c). The stochastic behaviour of the MBH orbit
was found to emerge when the ratio between the clump mass and black hole mass is greater than unity, as in this case
the impulsive interaction can modify the MBH orbit substantially. Decay time--scales are found to range from 1 to $\sim 50$ Myr, under the conditions explored.

Another type (c) investigation has been performed by \citet{delvalle15}, who recently simulated with an SPH code
the sinking of two MBHs in a massive $10^9\,\msun$ circumnuclear disc with gas forming stars. 
The orbits of the MBHs are erratically perturbed by the
gravitational interaction with the clumps that form as a result of disc's fragmentation, delaying the orbital decay of the MBHs  if compared with
similar runs with a one--component gas: 
typical decay times are found close to $10$ Myr, when the MBHs are seeded in the disc initially at 
$\sim 200-100$ pc scales. The key result which emerges from these new studies is that the MBH dynamics is sensitive not only to the time varying gravitational
background of a merger itself, but also on how fragmentation of gas clouds, star formation and supernova (SNa) feedback shape and change
the thermodynamical state of the gas, considered to play a key role in guiding the orbital decay of the MBH.
The mass distribution of the star--forming clumps appears to be a relevant parameter which affects the degree of stochastic forcing
of the MBH orbit and the distribution of the sinking times from $\sim 100$ pc scale  down to $0.1$ pc.

In this paper we simulate  the evolution of two gaseous discs [type (b)] and of their embedded black holes  
in the aim at studying the MBH dynamics within a multiphase gas shaped by cooling, star formation and SNa feedback, using the adaptive mesh refinement 
code (AMR) \textsc{ramses} \citep{teyssier02} modified to accurately track the MBH dynamics as presented in \citet{lupi15}.
AMR codes like \textsc{ramses} are known to better resolve gas shocks with respect to SPH codes \citep{agertz07}, thus allowing a more accurate description of the gas dynamics when the two gaseous discs collide.
A key question to pose is: in a star--forming medium what is the role of SNa feedback in shaping the gas mass distribution around the MBHs?
The transit from the binary phase II to phase III of gravitational wave inspiral depends on the strength of gas--driven migration in a circumbinary disc 
surrounding the MBH binary \citep{cuadra09,shi12,roedig11,roedig12,delvalle12}. 
Here we first attempt to explore under which conditions a circumbinary disc forms around the two MBHs and how this depends on
the recipes adopted  to model the physics of star--forming regions.

The paper is organised as follows. In Section 2 we describe the numerical setup used to simulate the  massive gas discs embedded in a stellar bulge and
the recipes introduced to model star formation and SNa feedback.
In Section 3 we present the results focusing on both the MBH dynamics and the properties of the multiphase gas surrounding the MBH binary.
Section 4 contains our conclusions. 

\label{sec:Intro}

\section{Numerical setup and initial conditions}
\label{sec:ic}
We consider the late stages of a galaxy gas rich merger, in which both galaxies host an MBH surrounded by a circumnuclear disc. Full details of the initial conditions can be found in Paper I. Here we simply summarise the basic features. 

We initially set each of the two merging nuclei in dynamical equilibrium, assuming they are constituted by three different components:
\begin{itemize}
\item a stellar spherical structure (termed `nucleus' hereafter) described by an Hernquist profile \citep{hernquist90}, defined in spherical coordinates as
%%%%%%%%%%
\begin{equation}
\rho_b(r)=\frac{M_b}{2\pi}\frac{a}{r\left(r+a\right)^3},
\end{equation}
%%%%%%%%%%
where $\rho_b(r)$ is the density as a function of radius  $r$, $M_b=2\times 10^8 \msun$ the total nucleus mass, and $a=100$ pc the nucleus scale radius
containing 1/4 of $M_b$;
\item an exponential gaseous disc with surface density profile defined (in cylindrical coordinates) as 
%%%%%%%%%%
\begin{equation}
\Sigma_d(R)=\frac{M_d}{2\pi R_d^2} \exp(-R/R_d),
\end{equation}
%%%%%%%%%%
where $R$ is the disc radius, $R_d= 50$ pc the disc scale radius containing 0.26 of the total disc mass $M_d=10^8 \msun$;
\item an MBH with mass $M_{\rm BH}=10^7 \msun$, at rest in the centre of the disc.
\end{itemize}

We build the two equal mass corotating gaseous discs, each described by $10^5$ particles, by means of the publicly available code \textsc{gd\_basic} \citep[see][for the algorithm description]{lupi15} and we relaxed them for about 10 Myr to ensure stability.  
The discs are initially set at $300$ pc on an elliptical orbit with eccentricity $e=0.3$, and with orbital angular momentum {\it antiparallel} to the angular momentum of the  discs. 
We stress that each galaxy disc plane is in
principle uncorrelated to the orbital plane of the merger, and, to the
first order, the same is valid for the CNDs\footnote{Here we are
neglecting the possible tidal effect exerted by one disc on to the other.
This effect would tend to align (or antialign) the two discs, enhancing
the chances of having an orientation between the two CNDs similar to the
one we assumed as initial conditions.}. We arbitrarily chose the geometry
that maximises the impact of the two discs along their orbit
and that ensures the highest cancellation of angular momentum,
enhancing the inflows towards the centremost regions. Such a geometry has
not been explored in the literature yet.
The initial conditions for the AMR runs are obtained by mapping the gas particle distribution on the grid using the publicly available code \textsc{tipgrid}.\footnote{The code is available at \url{http://www.astrosim.net/code/doku.php?id=home:code:analysistools:misctools}.}

We perform a total of six simulations using the AMR code \textsc{ramses} \citep{teyssier02}, where we change the
prescriptions regarding gas cooling, stellar mass formation (SF) and
SNa feedback to survey how the implementation of sub--grid
physics affects the evolution of the system.  The gas has primordial
composition, it is optically thin and cools down under lines and
continuum emission.  The maximum spatial resolution (at the highest
refinement level) for all our simulations is $\sim 0.39$ pc and the
mass resolution for particles forming the stabilising stellar nucleus is $2\times 10^3 \msun$.  The Jeans length is 
always resolved with at least 4 cells (14 in the
highest refinement level) and we also add a pressure support
  term, modelled as a polytrope with $\gamma=5/3$ and temperature
  $2\times 10^3$ K at the star formation threshold, in order to avoid
  the formation of spurious clumps due to resolution limits. 
  
In order to achieve the best possible treatment of MBH dynamics, we
adopt the additional refinement criterion described in
\citet{lupi15}.  We allow stellar particle creation when gas matches two criteria: (i) the gas temperature drops
  below $2\times 10^4$ K, and ii), the gas density in a cell
  exceeds a pre-defined value. We assume two different fiducial values for the SF
density threshold, i.e., $n_{\rm H}=2\times 10^5$ and $n_{\rm H}=2\times
10^6$ cm$^{-3}$, where $n_{\rm H}$ is the local hydrogen number
density, and a typical
SF time--scale of 1.0 Myr.  The resulting average mass of stellar particles is of $\sim
  300\,\msun$. Such value is significantly more massive than, e.g., what employed in \citet{amaroseoane13}, who however simulated a lighter and more compact system. We checked that our prescription
results in a gas--to--stellar mass conversion rate not lower than 
the local Kennicutt-Schmidt law. 

In  order to model SNa explosions, we consider each
stellar particle as a stellar population following a Salpeter IMF, and a SNa yield of 15\%. 
We further employ two different recipes for the SNa thermal
feedback. In both SNa feedback recipes the energy budget associated
($10^{50}\rm\, erg/\msun$) is completely released in the parent cell
as purely thermal energy. The first criterion (termed `ThFB' after
thermal feedback) assumes that the heated gas starts cooling right
after the SNa event; the second criterion (termed `BWFB' after blast
wave feedback) assumes instead that the energy released by SNe is
decoupled from the gas radiative cooling, i.e., it is not radiated
away for $\sim 20$ Myr \citep{teyssier13} and this triggers the
formation of a momentum--driven blast wave.  This latter scheme is
aimed at modelling non--thermal processes energising the blast wave,
which are characterised by time--scales longer than thermal processes
\citep[see e.g.][]{ensslin07}.  We usually assume that no star
formation occurs within the two discs before the merger, and that SNe
explode after a time $\Delta t_{\rm SN}=10$ Myr. Stellar mass particles
forming the stabilising bulge are not allowed to release energy as
SNe.

We run two further simulations at the highest density threshold for
star formation (termed `ThFBh\_prompt' and `BWFBh\_prompt') assuming
no time--lag between star formation and SN explosion. The aim of these
runs is to test the effects on the global (gas and BHs) dynamics of a
maximally fast SN feedback, comparing the results to the standard
$\Delta t_{\rm S}=10$ Myr case. While simulations with standard delay
are meant to model star formation as triggered by the merger of the
two circumnuclear discs, the 0--lag case may represent a situation
where sustained star formation is already in progress at the time of
the merger.

Finally, in order to avoid inaccurate integration of the orbits, all
runs are stopped when the MBH separation is approximatively three to four
times the cell length. Table \ref{tab:runs} summarises our six
simulations with the parameter used.

%%%%%%%%%%%%%%%%%%%%
\begin{table}
\centering
\begin{tabular}{llrl}
  Run &  $n_{\rm H}$ & $\Delta t_{\rm SN}$ & Feedback\\
~& (cm$^{-3}$) & (Myr)&\\
\hline
\hline
ThFBl &  $2\times 10^5$ & 10.0& Thermal\\
ThFBh &  $2\times 10^6$ & 10.0 & Thermal\\
BWFBl &  $2\times 10^5$ & 10.0 & Blast wave\\
BWFBh &  $2\times 10^6$ & 10.0 & Blast wave\\
\hline
ThFBh\_prompt &  $2\times 10^6$ & 0.0 & Thermal\\
BWFBh\_prompt &  $2\times 10^6$ & 0.0& Blast wave\\
\hline
\hline
\end{tabular}
\caption{{The complete suite of runs. The second column shows the density threshold for SF, the third column 
the lifetime of massive stars and the fourth column the type of feedback employed.}}
\label{tab:runs}
\end{table}
%%%%%%%%%%%%%%%%%%%%

\section{Results}
 
\subsection{Black hole dynamics}
\label{sec:dynamics}

We start by describing the MBH dynamics for the two simulations characterised by standard thermal SNa feedback and a typical time for SNa explosions of $10$ Myr (runs  `ThFBl' and `ThFBh'). These two runs are meant to represent a case where star formation is indeed triggered by the merger event, while gas thermodynamics is governed by standard thermal processes. The two different 
density thresholds for SF are used to assess the effects that  the efficiency of gas conversion into stars has on the MBH dynamical evolution. 

\begin{figure*}
\centering
\includegraphics[scale=0.44]{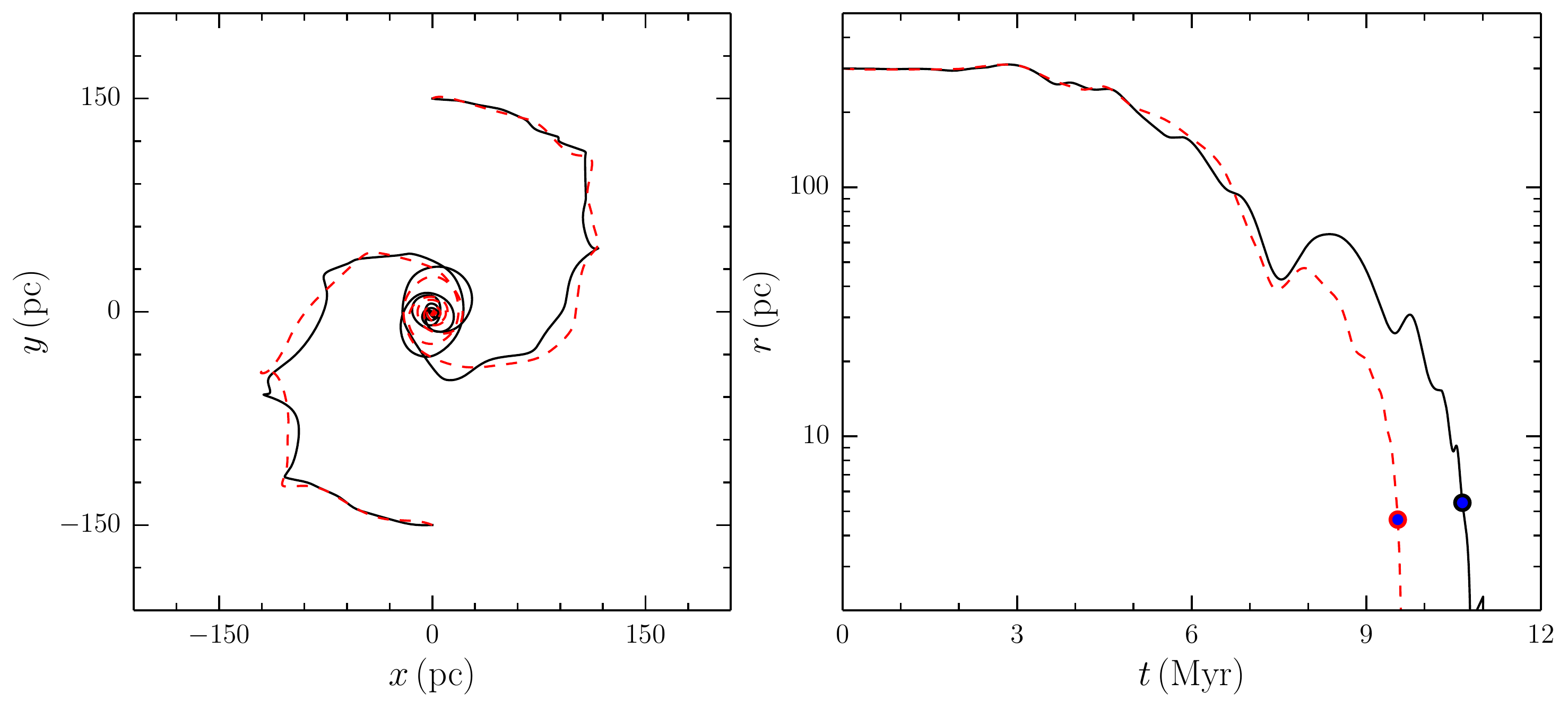}
\caption{{MBH dynamical evolution for runs `ThFBl' (solid black lines) and `ThFBh' (dashed red lines). Left--hand panel: projected orbital evolution. Right--hand panel: MBH separation versus  time. The blue dots correspond to the time of binary formation.}}
\label{fig:threshThermal}
\end{figure*}

In Fig. ~\ref{fig:threshThermal} we show the MBH projected orbits (left--hand panel), and MBH separation versus time (right--hand panel).  
The two MBHs exhibit a peculiar orbital motion, which can be explained when considering the gravitational interactions between the MBHs and massive gas/star clumps forming in the merging discs. Such interactions typically accelerate the orbital decay of the MBHs, and a gravitationally bound MBHB
forms after $\sim 10$ Myr (the binary formation time is indicated as a blue dot in the right--hand panel). 
In the case of ThFBh run (dashed red lines), the MBH orbits appear more perturbed, and the orbital decay is somewhat faster. 

\begin{figure*}
\includegraphics[width=0.75\textwidth]{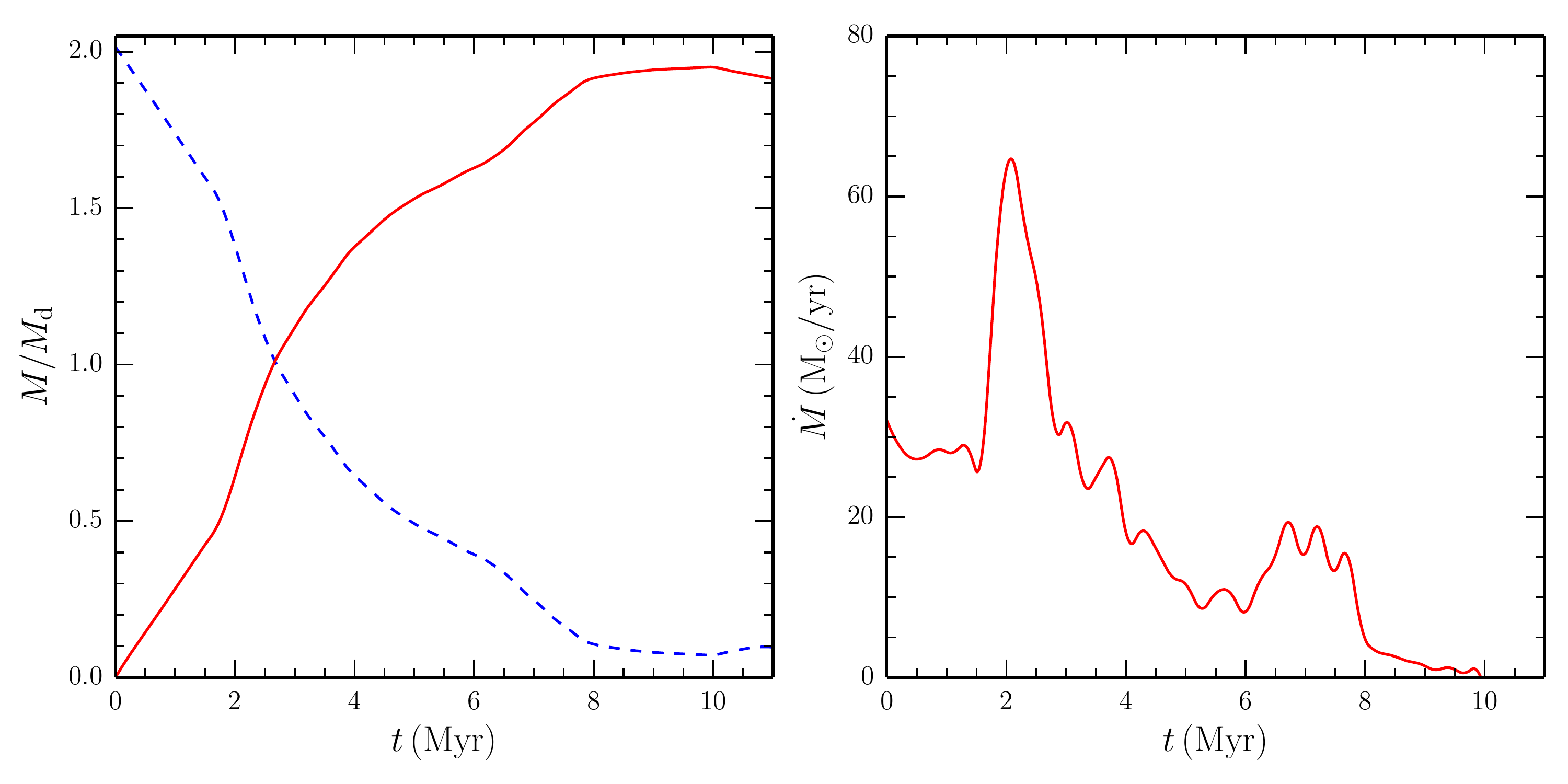}
\caption{{Star formation in run `ThFBl'. Left--hand panel: total stellar mass (solid red line) in units of the initial disc mass $M_d$, and the residual gas mass 
in units of $M_d$ (dashed blue line)  as a function of  time. Right--hand panel: star formation rate versus time.}}
\label{fig:sfrate}
\end{figure*}

\begin{figure}
\includegraphics[scale=0.408]{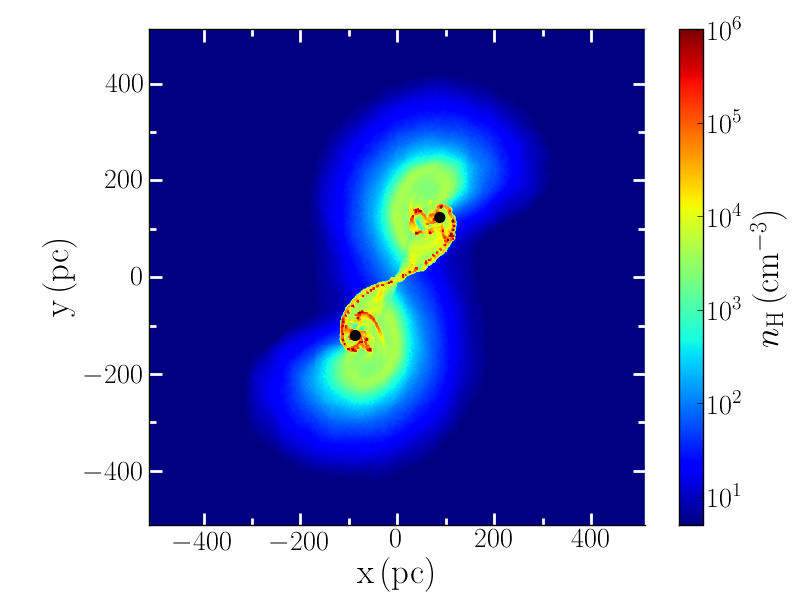}
\caption{{Face--on gas density map for run `ThFBl' at time $t=2.1$ Myr (i.e., when the star formation rate is maximum, see Fig.~\ref{fig:sfrate}). The gas shocked after the first disc collision fragments into a large number of small clumps which very rapidly convert gas into new stellar mass. The black dots correspond to the positions of the two MBHs.}}
\label{fig:gas2.1}
\end{figure}

Fragmentation of gas, occurring just after the simulation starts, tends to form massive gas clumps, especially in the high--density regions surrounding the two MBHs. In high--density clumps star formation is very effective, and overall, a large fraction of the initial disc gas is converted into stellar mass within $~10$ Myr. This is apparent from Fig. ~\ref{fig:sfrate} (left--hand panel), where the stellar mass and the residual gas mass are shown as a function of time. The right--hand panel of Fig. ~\ref{fig:sfrate}, instead, shows the star formation 
rate versus time. A fast increase of star formation occurs initially since gas shocked during the disc collision fragments into small clumps which immediately convert into stellar particles. After $\sim 2$ Myr, only low--density gas survives. Hence, star formation is no longer efficient and almost steadily decreases in time. In Fig.~\ref{fig:gas2.1} we plot the mass--weighted gas density map at time $t=2.1$ Myr, defined as the time of the peak in star formation rate (see Fig.~\ref{fig:sfrate}, right--hand panel). 

\begin{figure}
\centering
\includegraphics[width=0.48\textwidth]{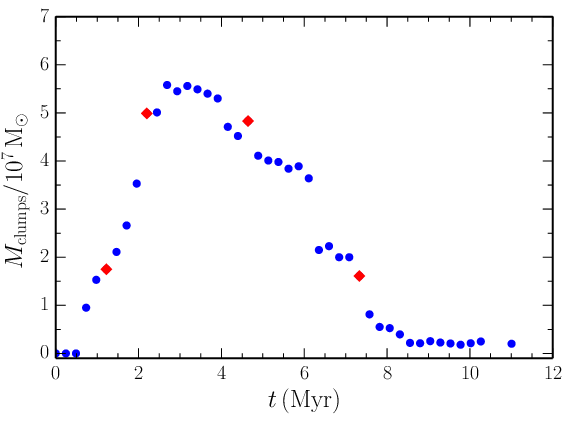}
\caption{{Total mass in clumps for run `ThFBl'. The red diamonds correspond to the times at which we computed the clump mass distribution shown in Fig.~\ref{fig:clmassdist}.}}
\label{fig:clumpmass}
\end{figure}

In order to quantify the impact of gas clumps on MBH dynamics, we estimate the total mass in gas/star clumps, along with the clump mass distribution. We define `clumps' those gravitationally bound regions that feature a single peak in the 3D density field. 
Fig.~\ref{fig:clumpmass} shows the total mass in clumps  as a function of time for run `ThFBl'. Two distinct phases can be observed, with a peak in the total mass of clumps occurring after a time $t_{\rm peak} \sim 2.5$ Myr. The initial fast growth of the gas locked in clumps is the result of the collision between the two unperturbed gaseous discs. Indeed, gas fragmentation is promoted along the shock surface (resulting also in the peak of star formation rate, see Fig.~\ref{fig:sfrate}, right--hand panel). 

\begin{figure*}
\includegraphics[scale=0.64]{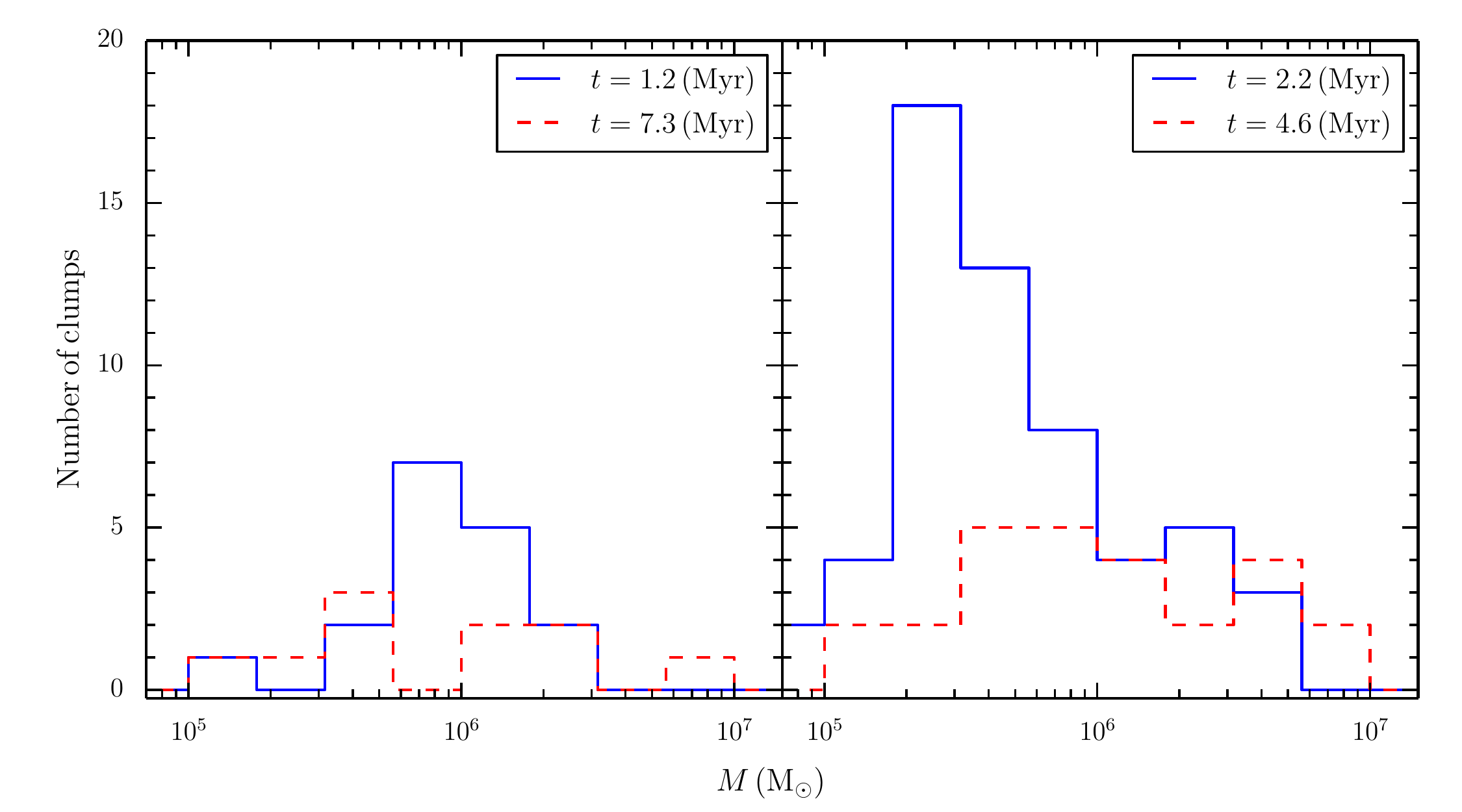}
\caption{{Mass distribution of clumps in run `ThFBl'. Left--hand panel: mass distribution at two selected times when the number of clumps is relatively small. Right--hand panel: same as left--hand panel, but at two times when the number of clumps is larger. The four selected times are marked as red dots in Fig.~\ref{fig:clumpmass}.}}
\label{fig:clmassdist}
\end{figure*}

Fig.~\ref{fig:clmassdist} shows, for the same `ThFBl' run, the mass distribution of clumps at four different selected times marked as red dots in 
Fig.~\ref{fig:clumpmass}. We selected two times corresponding to a relatively low total clump mass ($\sim 1.8\times 10^7$ $M_\odot$, left--hand panel) and two times corresponding to a larger mass value ($\sim 5 \times 10^7$ $M_\odot$, right--hand panel), respectively, one before and one after $t_{\rm peak}$. The mass distribution lies in the range $10^{5-7}\,\rm M_\odot$, with few clumps as massive as the MBHs. These very massive clumps typically form after $t_{\rm peak}$, most probably due to gas accretion from low--density regions and to mergers between less massive clumps, and eventually will merge with the gas overdensity surrounding each MBHs. When one of these more massive clumps manages to approach an MBH at close range, then a transient MBH--clump binary system forms, strong gravitational perturbations develop, and the MBH orbit greatly deviates from its original path. This is the reason behind the `wiggling orbits' seen in Fig.~\ref{fig:threshThermal}, right--hand panel. 
The typical BH--clump distance when the transient binary system forms is $\sim 10-20$ pc, which is always resolved with a number of cells $\gsim 10$, thanks to the refinement prescription described in \citet{lupi15}, allowing us to accurately resolve the BH--clump close interaction.

In the case of run `ThFBh', because of the relatively higher density threshold for SF, a slightly larger number of more massive clumps forms, resulting in the more disturbed orbits (and faster decay) seen in Fig.~\ref{fig:threshThermal}. 

%%%%%%%%%%%%%%%%%%

%%%%%%%%%%%%%%%%%%%

We have then compared the above analysis regarding the MBH dynamics with runs employing the aforementioned blast wave--like feedback from SNe (BWFB--like runs). As discussed before, this feedback implementation aims at describing non--thermal processes in the aftermath of SNa explosions. We find that the dynamical evolution of the MBHs is largely independent upon the details of the SNa feedback employed, making MBH dynamics results fairly robust against the different implementations of sub--grid physics.

\subsection{Gas dynamics}

We discuss here the dynamics of the gas during the merger event. We focus on the case with the low--density threshold for SF (run `ThFBl'), keeping in mind that the higher density case produces a qualitatively and quantitatively similar outcome.

\begin{figure*}
\centering
\includegraphics[scale=0.395]{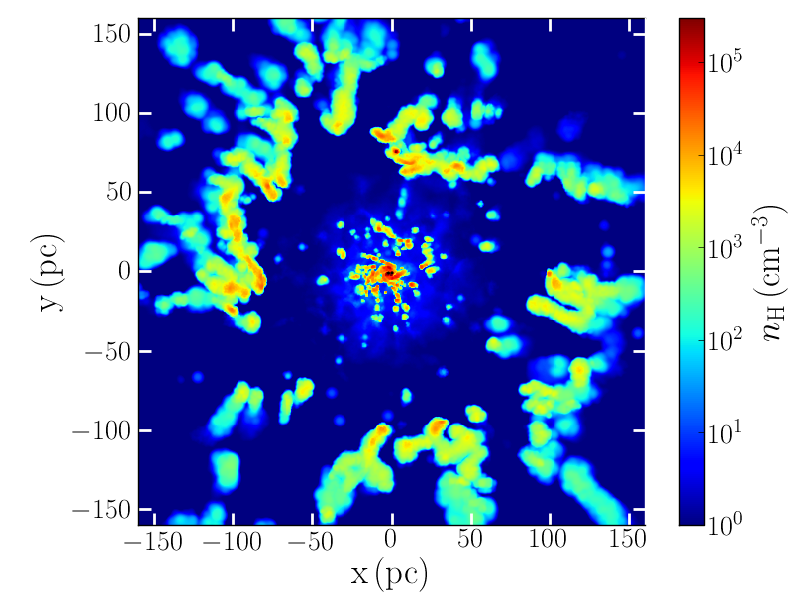} 
\includegraphics[scale=0.395]{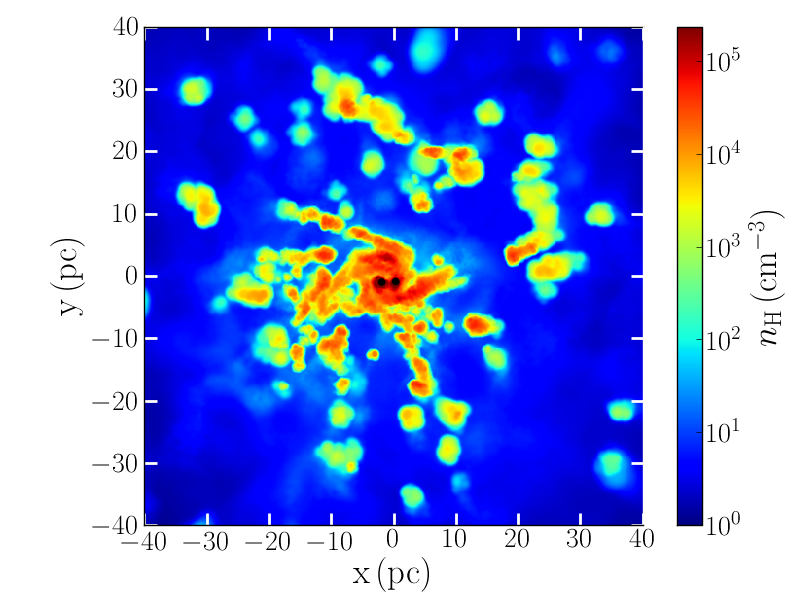}
\caption{{Face--on gas density maps for run `ThFBl' around the MBHB at the end of the simulation ($t\sim 11$ Myr). Left--hand panel: on large scales the disc is almost totally disrupted because of SNa explosions. Right--hand panel: zoom in of the nuclear region where an inner corotating gas disc forms.}}
\label{fig:discThFBl}
\end{figure*}

Fig.~\ref{fig:discThFBl} shows the gas distribution around the MBHB
after $t=11$ Myr. On large scale (left--hand panel), the relic disc
resulting from the collision of the progenitor discs is almost totally
disrupted because of SNa feedback. This residual structure is
counter--rotating relative to the MBHB orbit.  On scales of the order of
few pc (right--hand panel), the gas which has not been converted into
stellar particles settles in a circumbinary disc, with a total mass
of few $10^5\rm\, M_\odot$.  The small disc corotates with the MBHB
thanks to the dragging of gas by the MBHs during their inspiral towards
the centre. Note that this implies that the angular momentum of the
residual gas changed sign during the evolution of the system.

\begin{figure}
\centering
\includegraphics[width=0.47\textwidth]{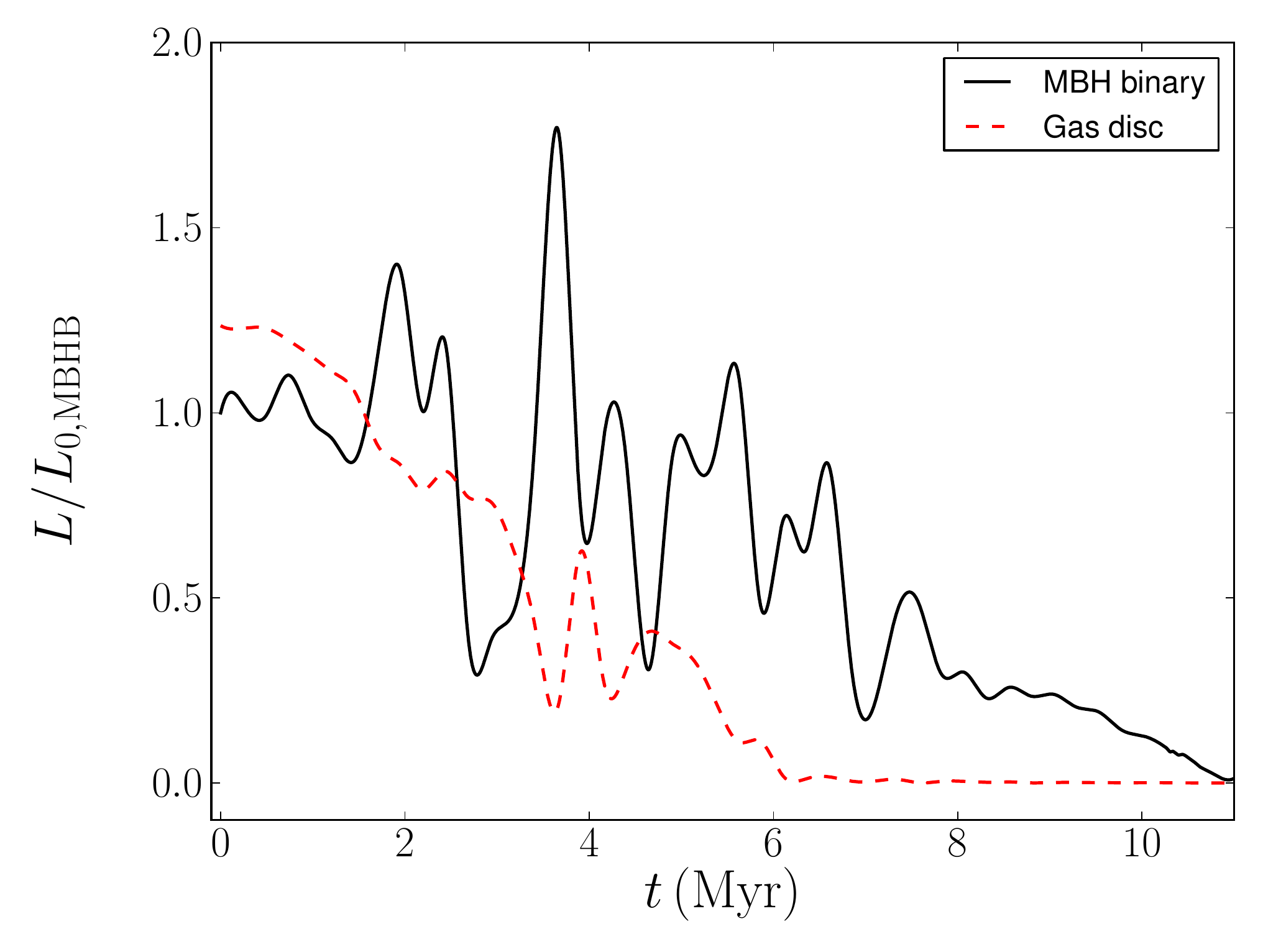}
\caption{{Evolution of the moduli of the angular momenta  for run `ThFBl'. The MBHB  total angular momentum (solid black line) is compared with the gas angular momentum (dashed red line), defined as the total angular momentum of the gas within a sphere centred on one MBH whose radius equals half the MBHB separation. Both curves have been normalised to the initial angular momentum of the MBHB.}}
\label{fig:angularevo}
\end{figure}

We report in Fig.~\ref{fig:angularevo} the evolution of the modulus
of MBH orbital angular momentum and compared it to the modulus of the
total angular momentum of the gas which is the closest to the MBHs in
the simulation, defined as the gas within a sphere of radius equal to
0.5 times the MBH separation. We observe that at the beginning of the
simulation the angular momentum of the gas is larger than that of the
MBHs, and we remind that the gas is counter--rotating.  After $\gsim
4$ Myr, the angular momentum associated with the MBH orbit exceeds that
of the gas and in principle there are the conditions for a change in
the sign of the gas angular momentum, being dragged by the MBHs.  The
gas angular momentum actually changes sign after $\sim 9$ Myr, when
the MBH separation is $\sim 45$ pc. At this evolutionary stage, a
large fraction ($\gsim 90\%$) of the initial gas mass is already
converted in stellar particles. After $\simeq 10$ Myr, when SNe start
to explode, the released energy is radiated away by the small amount
of residual gas, which is however unable to form further stellar mass
at a comparable rate. In other words, star formation is not halted by
SNa feedback, rather by gas consumption.

Concerning the impact of blast wave feedback (BWFB--type runs), as
expected it does not alter the gas dynamics for a time $\sim \Delta
t_{\rm SN}$ (at that point the two MBHs have already reached the
centre of the system). After that time, the almost simultaneous SNa
events release a fairly large amount of energy which heats the gas up
but is not radiated away. The net result is that the remaining gas is
pushed at very large distances from the MBHB (up to $\sim 500$ pc) by
the increased pressure. The MBHB lives then in a very low--density
environment, and no circumbinary disc is formed on any scale.

%%%%%%%%%%%%%%%%%%%
%%%%%%%%%%%%%%%%%%%

%%%%%%%%%%%%%%%%%%%%%%%%%%%%

\begin{figure*}
\centering
\includegraphics[scale=0.50]{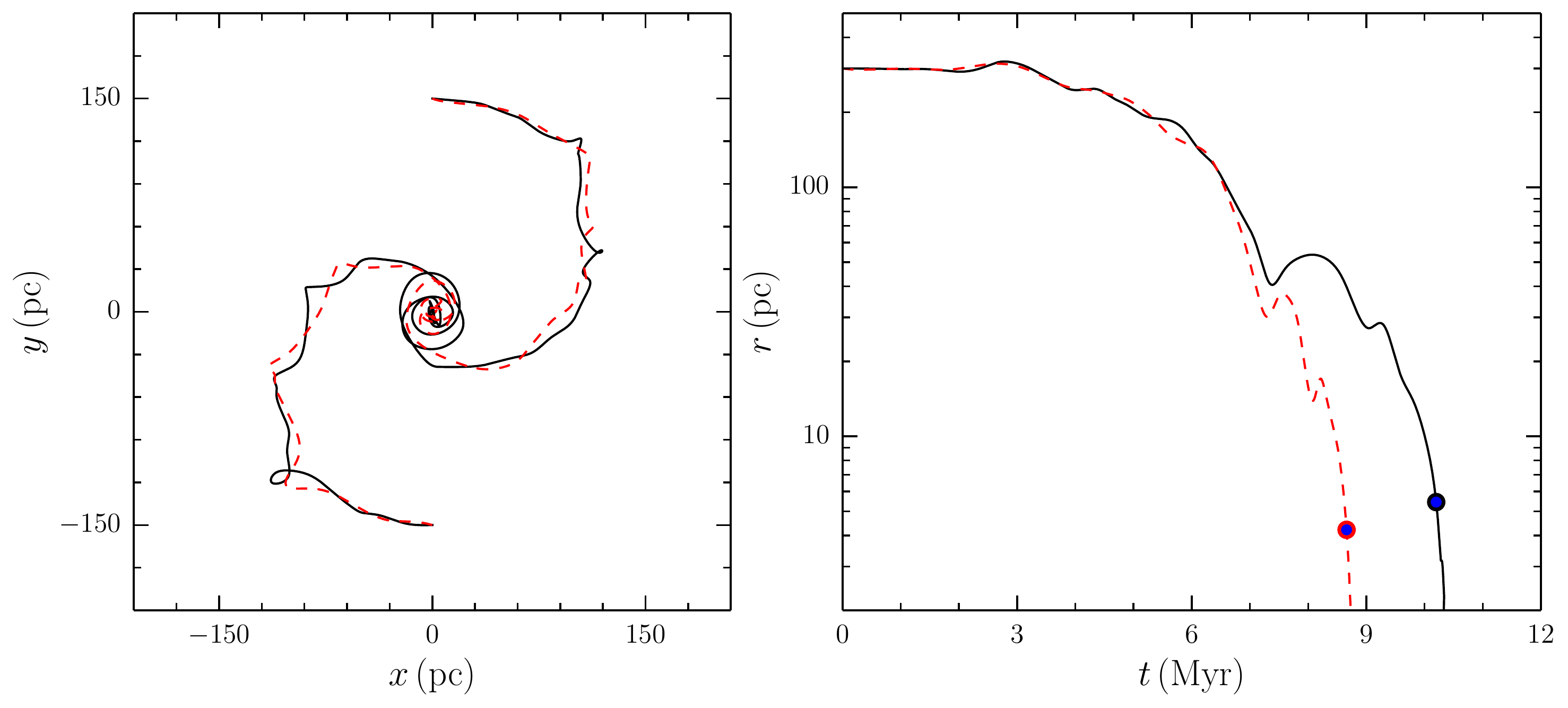}
\caption{{Same as Fig.~\ref{fig:threshThermal} but for runs `BWFBl' (solid black lines) and `BWFBh' (dashed red lines).}}
\label{fig:threshBlast}
\end{figure*}

\subsection{Prompt SNa explosions}

Both the MBH and gas dynamics are unaffected by feedback for the first 10 Myr as this is the assumed lifetime of massive stars (and hence for the onset of SNa feedback). To test how our results depend upon such choice, we 
cosnider the extreme case of 
$\Delta t_{\rm SN}=0$ Myr, i.e, massive stars explode as soon as they form. 

We find that, as long as the SNa feedback is governed by thermal processes, only small differences in the MBH dynamics exist 
compared to the standard delay case previously discussed. This similarity occurs because the SNa energy is mostly released in high--density clumps, where gas cools down very rapidly, and the clumps can survive the explosion. As a consequence, star formation can proceed  
until almost all clump gas is consumed.  

\begin{figure*}
\centering
\includegraphics[scale=0.50]{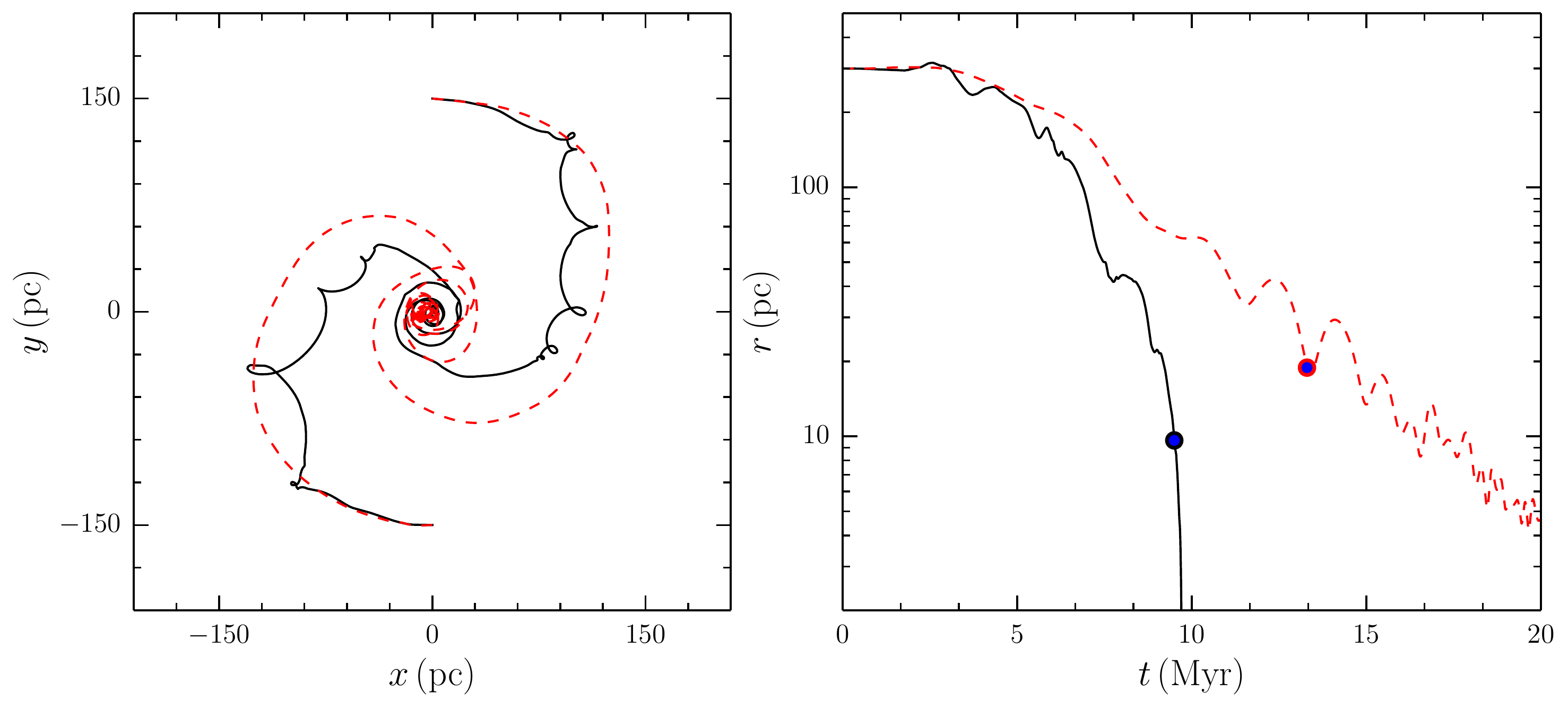}
\caption{{Same as Fig.~\ref{fig:threshThermal} but for runs `ThFBh\_prompt' (solid black lines) and `BWFBh\_prompt' (dashed red lines). The blue dots correspond to the time of binary formation.}}
\label{fig:threshBWFB_prompt}
\end{figure*}

Large differences occur instead if, along the $\Delta t_{\rm SN}=0$ assumption, we employ the blast wave recipe for SNa feedback. 
In Fig.~\ref{fig:threshBWFB_prompt} we compare the projected MBH orbits (left--hand panel) and the MBH separation versus time (right--hand panel) for runs BWFBh\_prompt and ThFBh\_prompt.  
In the case of blast wave like feedback, the orbital decay is slower, with a typical binary formation time--scale of $\gsim 13$ Myr. The difference is due to the early SNa explosions that, coupled with the blast wave--like feedback, tend to disrupt the gas clumps and to deplete the gas reservoir progressively forming around the MBHs. As a consequence, the two MBHs evolve in a lower density, smoother environment, where low--mass clumps are typically unable to induce strong orbital perturbations. The net result is a less disturbed orbital decay (Fig.~\ref{fig:threshBWFB_prompt}, left--hand panel). 

We therefore conclude that in the case of prompt SNa explosions, contrary to the standard delay case, the dynamical evolution of the MBHs is strongly affected by the feedback mechanism  employed. The SF density threshold instead does not result in relevant differences anyway. 

\begin{figure*}
\centering
\includegraphics[scale=0.395]{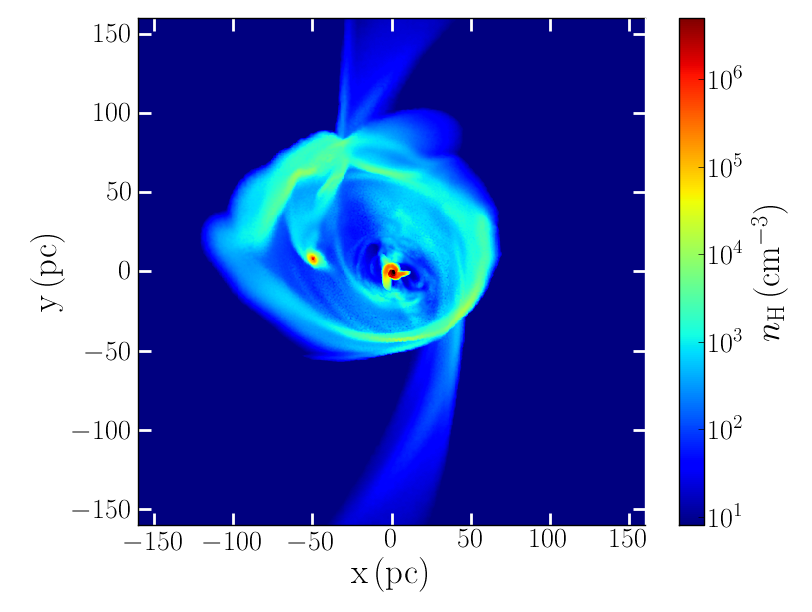}
\includegraphics[scale=0.395]{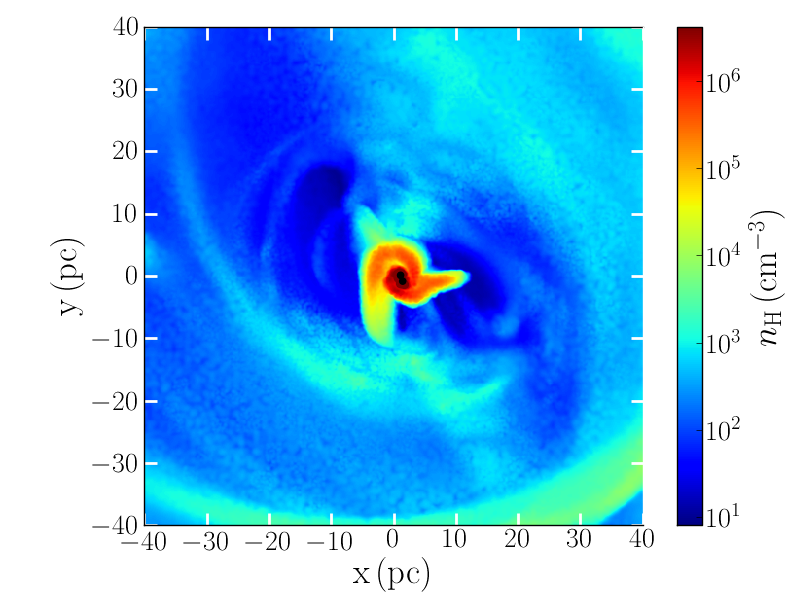}
\caption{{Same as Fig.~\ref{fig:discThFBl} but for run `ThFBh\_prompt' at time $t\sim 10$ Myr. Left--hand panel: gas settles in a disc/ring like structure
which is counter--rotating relative to the MBHs. Right--hand panel: zoom in of the  region where an inner corotating gas disc forms around the  
MBHB visible on the east side of the left--hand panel.}}
\label{fig:discThFBh_prompt}
\end{figure*}

While MBH dynamics is basically unaffected by the value of $\Delta t_{\rm SN}$ in the case of thermal SNa feedback, substantial differences occur 
in the dynamics of the gas component. Along with a small--scale corotating circumbinary disc, we do observe a further, much larger disc/ring--like structure on $\sim 100$ pc scale (see Fig~\ref{fig:discThFBh_prompt}). Indeed, feedback from SNe does not occur suddenly after 10 Myr but it is instead diluted in time, so that the (rapidly cooling) gas has time to readjust in a disc--like structure.  Though several other possible explanations exist (e.g., secular evolution of the Galactic disc), it is tempting to associate such structure to the central molecular zone of the Milky Way \citep{jones11}. It is interesting to note that the larger scale disc keeps memory of the initial angular momentum, and it is then counter--rotating with respect to the small inner circumbinary disc which 
is, as discussed above, dragged by the MBHB. 

\begin{figure*}
\centering
\includegraphics[scale=0.3]{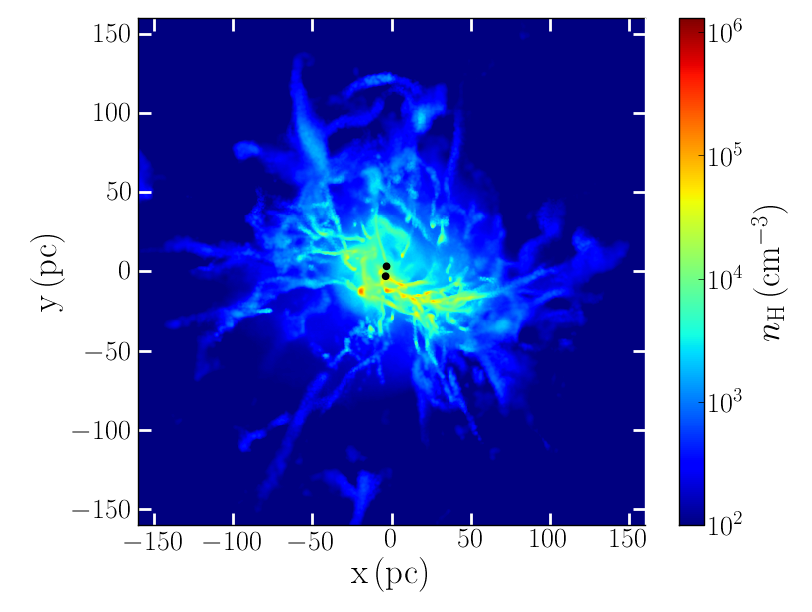}\\
\includegraphics[scale=0.3]{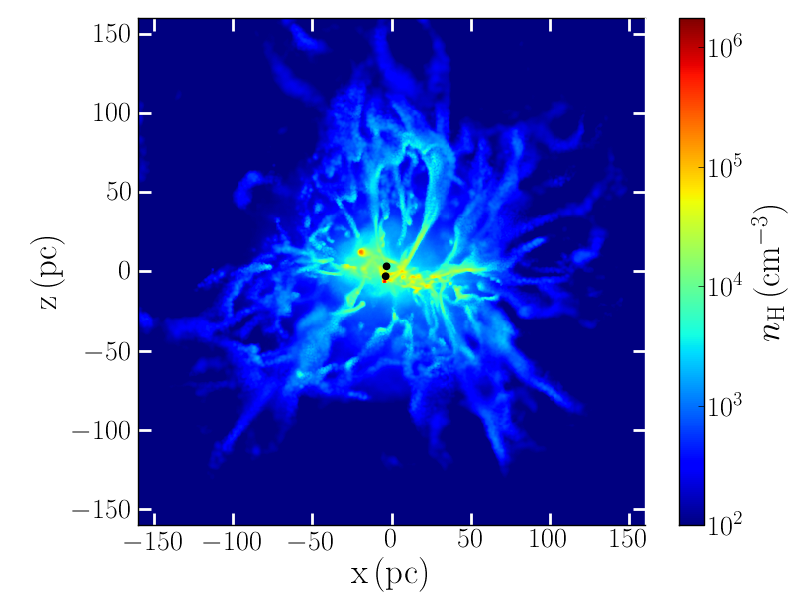}
\includegraphics[scale=0.3]{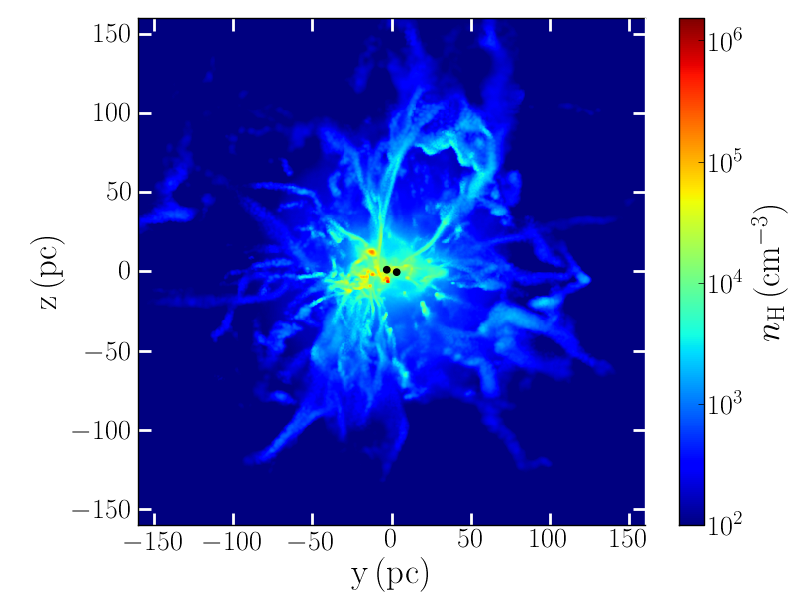}
\caption{{Gas density maps for run `BWFBh\_prompt' around the MBHB  at the end of the simulation ($t\sim 20$ Myr). The gas settles in a triaxial structure with a denser central core. The core mass is $\sim 10^7\rm\, M_\odot$ within a radius $\sim 25$ pc. The upper panel shows the face--on view, while the two bottom panels show the edge--on views
of the triaxial gas configuration.}}
\label{fig:triaxBlFBh_prompt}
\end{figure*}

The case of blast wave--like feedback is still different. We do not observe a disc--like structure, rather we find a massive triaxial gas distribution surrounding the MBHB with density of  few $10^5$ cm$^{-3}$ (see Fig.~\ref{fig:triaxBlFBh_prompt}).
This difference is produced by the different nature of the SNa feedback, which is in this case able to heat the gas and provide a pressure support  large enough to prevent gas contraction.

Because of the large fraction of gas available (due to the SNa feedback which reduces the net star formation by destroying gas clumps, as discussed above) the gas will continue to cool down, resulting in alternated phases of star formation (due to gas cooling and contraction) and re--heating (due to SNa feedback). We observe a large number of dense gas streams flowing from low--density regions towards the centre where the MBHB resides. This large inflow will result in a burst of star formation in the nucleus and in a following phase of SN explosions. The energy provided by SNe will then reheat the gas, stopping the contraction and eventually expand the entire gas structure into a less dense state. These alternated phases, if occurring for enough time, could convert a large fraction of gas into new stellar mass, which could eventually form a massive nuclear stellar cluster surrounding the MBHB.

%%%%%%%%%%%%%%%%%

\section{Discussion and Conclusions}
By means of high--resolution,
AMR hydrodynamical simulations, 
we explored the evolution of two massive gas discs hosting at their centre an MBH. The two discs are on an elliptic orbit and merge, 
to mimic the encounter between two very gas--rich disc galaxies.   To maximise the strength of the 
interaction, the orbital angular momentum is chosen to be antiparallel to the disc's angular momenta.

Strong shocks that develop during the
merger  of the two discs become  sites of intense star formation, and stellar feedback alters significantly the thermal and dynamical state of the gas
which undergoes a major transformation. Most of the gas is turned into new stellar particles  through the formation of clumps 
of mass $\lesssim 10^6\msun$. Only few clumps form as massive as the MBHs, weighing $10^7\msun$. 

We explored different SNa feedback recipes: the thermal and blast wave feedback, assuming a lifetime of $\sim 10$ Myr for the massive stars.
Furthermore, we considered a case  in which  prompt SNa explosion is coupled with both thermal and blast wave feedback.
We find that the orbits of the two MBHs are perturbed due to their interaction with single clumps during the paring phase I, resulting in impulsive kicks that 
imprint sudden changes in the direction and velocity of the orbit.
Sinking times of  $\sim 10-20$ Myr are found, considering the set of parameters used.  The pairing phase terminates with the
formation of a Keplerian binary.

The MBH orbit is stochastic due to the presence of  the gas clumps. However, we do not
see a sizeable delay or spreading in the sinking time due to gas clumpiness, contrary to what found in \citet{fiacconi13}, where
the level of stochasticity of the orbit was higher.
We interpret this difference as due to the geometry of
the collision that mainly confines star formation along the oblique shock forming at the time of impact of the two discs, and to the fact that 
in our case the mass distribution of the clumps evolves as gas is turned into stellar mass which spread due to dynamical relaxation.
Our simulated MBHs do not leave the orbital plane due to clump--induced kick, contrary to what seen in \citet{roskar15}, as our simulation is strictly coplanar.
We expect that an inclined encounter would lead to a change in the orbital plane also in our case, and this will be explored
in future. During the pairing phase, the MBH dynamics is mostly affected by the presence of clumps and not by the recipe used to
model the feedback processes.

We note that, on the contrary, the gas distribution around the MBHs is
significantly affected by feedback.  Thermal feedback leaves no large--scale disc around the MBHB.  Yet a residual corotating
circumbinary disc of mass much smaller than the MBH mass forms
around the two black holes which we expect will control the further
spiral-in via migration--like mechanisms.

Blast wave feedback  is a way to model the expansion of SNa--driven bubbles.  With the code it is then possible to  mimic  the ballistic phase
of the shock triggered by the SNa explosion.
 As cooling is shut off in this phase, a multiphase gas forms and the sweeping of the gas induced by the blast wave
 leads to the almost complete evacuation of gas.  The MBHB thus inhabits a region completely devoid of gas.
 Blast wave feedback in the prompt scenario leads instead to a configuration in which the MBHB is surrounded 
  by a gas cloud with little angular momentum and triaxial in shape.

The lesson to learn is that star formation in merging gaseous discs is
a key process which affects the physical state of the gas in the
surroundings of the MBHs. Under these circumstances it is difficult to predict the actual
distribution of gas when the most active phase of the merger has
subsided, as the outcome depends upon the modelling and on sub-grid physics, and firm conclusions should be taken with caution. Still, the presence of cool gas has deep implications for
  the evolution and observability of close MBHBs. First, the evolution
  of a binary on sub-pc scales towards the coalescence is strongly
  dependent on the gaseous and stellar distribution in its immediate
  surroundings \citep[][for a review]{colpi11}. The time--scale of the MBHs
  shrinking on sub--pc scales is of fundamental importance as it
  affects the expected rate of binaries possibly observable as gravitational
  wave sources. This is particularly true in mergers between gas--rich
  galaxies, a fraction of which can host binaries detectable by future
  space--based gravitational wave detectors such as $eLISA$
  \citep{amaro-seoane13}. Secondly, the
  presence of gas is a necessary condition for the possible detection
  of the binary during the hardening phase \citep{dotti12} as
  well as for pinpointing an electromagnetic counterpart of the MBHB coalescence
  \citep[see e.g.][]{schnittman13,bogdanovic15}.  The lack of a
  clear consensus on the processes shaping the environment of MBHBs, whose
  evolution actually depends on the physical modelling, and the lack of 
  observations available on the small scales we considered in our work 
  witness the need of investigating a wider range of parameters.

\section{Acknowledgements}
We thank R. Teyssier and L. Paredi for many fruitful discussions. We acknowledge financial support from italian MIUR, through PRIN 2010-2011. Simulations were run on the EURORA cluster at CINECA and on the Lucia cluster at DiSAT, University of Insubria.
\bibliographystyle{mn2e} 
\bibliography{MBHB}
\label{lastpage}
\end{document}